\documentclass[fleqn,usenatbib]{mnras}

\usepackage{newtxtext,newtxmath}

\usepackage[T1]{fontenc}
\usepackage{ae,aecompl}

\usepackage{graphicx}	
\usepackage{amsmath}	
\usepackage{amssymb}	

\usepackage[normalem]{ulem}

\title[Galaxy rotation helps SMBHs merge]{Galaxy Rotation and Supermassive Black Hole Binary Evolution}

\author[M. A. Mirza et al.]{M. A. Mirza,$^{1}$ \thanks{E-mail: awaismirza786@yahoo.com}
A. Tahir,$^{1}$
F. M. Khan,$^{1\thanks{E-mail: khanfazeel.ist@gmail.com}}$
K. H. Bockelmann,$^{2,3}$
A. M. Baig,$^{1}$
\newauthor
P. Berczik$^{4,5,6}$
and F. Chishtie$^{7}$
\\
$^{1}$Department of Space Science, Institute of Space Technology, Islamabad 44000, Pakistan\\
$^{2}$Department of Physics and Astronomy, Vanderbilt University, Nashville, TN 37240, USA\\
$^{3}$Department of Physics, Fisk University, Nashville, TN 37208, USA\\
$^{4}$The International Center of Future Science of the Jilin University, 2699 Qianjin St., 130012, Changchun City, China\\
$^{5}$National Astronomical Observatories of China and Key Laboratory for Computational Astrophysics, Chinese Academy of Sciences, \\
20A Datun Rd, Chaoyang District, 100012 Beijing, China \\
$^{6}$Main Astronomical Observatory, National Academy of Sciences of Ukraine,27 Akademika Zabolotnoho St, 03680 Kyiv, Ukraine \\
$^{7}$Theoretical Research Institute, Pakistan Academy of Sciences (TRIPAS), Islamabad 44000, Pakistan.
}

\date{Accepted XXX. Received YYY; in original form ZZZ}

\pubyear{2016}

\begin{document}
\label{firstpage}
\pagerange{\pageref{firstpage}--\pageref{lastpage}}
\maketitle

\begin{abstract}
Supermassive black hole (SMBH) binaries residing at the core of merging galaxies are recently found to be strongly affected by the rotation of their host galaxies. The highly eccentric orbits that form when the host is counterrotating emit strong bursts of gravitational waves that propel rapid SMBH binary coalescence.  Most prior work, however, focused on planar orbits and a uniform rotation profile, an unlikely interaction configuration. However, the coupling between rotation and SMBH binary evolution appears to be such a strong dynamical process that it warrants further investigation. This study uses direct N-body simulations to isolate the effect of galaxy rotation in more realistic interactions. In particular, we systematically vary the SMBH orbital plane with respect to the galaxy rotation axis, the radial extent of the rotating component, and the initial eccentricity of the SMBH binary orbit. We find that the initial orbital plane orientation and eccentricity alone can change the inspiral time by an order of magnitude. Because SMBH binary inspiral and merger is such a loud gravitational wave source, these studies are critical for the future gravitational wave detector, LISA, an ESA/NASA mission currently set to launch by 2034.
\end{abstract}

\begin{keywords}
black hole physics -- galaxies: kinematics and dynamics -- galaxies: nuclei -- rotational galaxies -- gravitational waves -- methods: numerical
\end{keywords}



\section{Introduction}\label{sec-intro}
With well over two decades of observational evidence, it is now understood that supermassive black holes (SMBHs) are present at the heart of nearly every galaxy \citep[][e.g.]{Ferrarese+05,Gultekin+09}, with masses in the range 10$^6$ -- 10$^{10}$ $M_\odot$\citep{Kormendy+13}.  Although the observational evidence is less iron-clad, galaxies are increasingly found with what appear to be dual AGN \citep{Komossa+06,Comerford+15} and even candidate SMBH binaries -- some separated by as little as 7 parsec \citep{Rodriguez+06}. Multiple SMBHs in a single galaxy are thought to mark a prior galaxy merger, because after a galaxy merger, SMBHs presumably sink to the center of the remnant and coalesce~\citep[][e.g.]{Begelman+80, Milosavljevi+01}.  The decay of a SMBH orbit is dominated by dynamical friction until a hard binary SMBH forms in the remnant core~\citep{Yu+02,Colpi+14}.  Once a hard binary forms, few-body scattering of stars by the SMBH binary extracts energy from the orbit, shrinking the SMBH separation \citep{Hills+83,Quinlan+96,Sesana+06} until gravitational radiation completes the coalescence \citep{Centrella+10}. A binary SMBH merger is expected to be the loudest gravitational wave source in the Universe; the merger of two 10$^6 \, M_\odot$ black holes, for example, would be detectable by LISA (Laser Interferometer Space Antenna) to redshift 3 with an SNR $>$ 100 \citep{goat+13}.

	It was quickly established that the scattering phase could sabotage this orderly progression to SMBH coalescence; when there are too few stars in the loss cone to drain energy from the binary orbit, the SMBHs are stranded at separations too large to coalesce in a Hubble time \citep{Begelman+80,Makino+04,berczik+05}. Although it was less quickly established, an emerging picture is that deviations from spherical symmetry, rotation, or an ample inflow of gas -- all natural outcomes of galaxy merging \citep{Khan+16,Gualandris+12} -- can efficiently usher the SMBH binary through the final parsec to coalesce \citep{Khan+11,preto+11,khan+13}. Work is shifting from {\it whether} SMBHs coalesce to understanding the mapping between galaxy properties and SMBH merger timescales, which ranges from tens of Myr to a few Gyr for simplified galaxy models and merger parameters \citep{khan+12b,vasiliev+15,sesana+15}.
	
	Now that we know that the SMBH host galaxy plays such a significant role in driving SMBH coalescence, it is important to build galaxy models that better reflect observations. Overwhelmingly, observations find that spiral bulges, S0s and ellipticals rotate~\citep[][e.g.]{Kormendy+04, Cappellari+07, Emsellem+11}, so it is clear that any study of SMBH coalescence in a non-rotating model is incomplete. 
    
    Recent work has consistently shown that a rotating stellar background alters the orbit of a SMBH binary. For example, in a cuspy spherical rotating model, a co-rotating SMBH binary orbit tends to circularize, while counterrotation pushes the orbit to high eccentricity \citep{Sesana+11}. Using a similar spherical stellar background potential, \citet{Gualandris+2012} focused on the orbital plane evolution of a massive SMBH binary with a high mass ratio, and noted considerable reorientation of the orbital plane to align with that of the underlying rotating host.

In a companion work, \citep{Holley-Bockelmann+15} (hereafter HB15) obtained that galaxy rotation speeds up SMBH coalescence when the binary is perfectly co- or counter-rotating with respect to its galaxy host. In the corotating case, the SMBH binary circularizes and settles into an orbit where the binary center of mass is in corotation resonance with the galaxy nucleus;  here, the affinity between stellar velocities and the binary center of mass renders dynamical friction less effective, while the center of mass motion and position facilitate more stellar interactions. On the other hand, counterrotation pumps up the eccentricity of the SMBH binary  \citep[see also][]{Sesana+11,Wang+14}, leading to rapid coalescence as the close pericenter passes emit large surges of gravitational radiation that drain energy from the binary orbit. Such sources of gravitational waves create ripples in space-time with signal to noise ratios loud enough to be detectable out to redshift 20. Binary black hole mergers are the prime target of future space based detectors, such as LISA (Laser Interferometer Space Antenna)\citep{elisaConsortium+13}, as probes of cosmology, black hole growth and galaxy evolution.
	
		While it is clear that bulk galaxy rotation is, in principle, a particularly effective way to drive SMBH mergers, prior work was limited to highly simplified configurations; it is not clear how these results 
extrapolate to more generic cases of binary orbital plane misalignment, partial rotation, and various initial eccentricities.  In this paper, we study SMBH binary coalescence as a function of galaxy rotation, initial binary orbital plane orientation, and initial binary eccentricity. Our setup and study extends \cite{Sesana+11} and \cite{Gualandris+2012} by using flat rotating cuspy galaxy models rather than spherical models. Flattened galaxy models are capable of driving SMBH binary evolution all the way to gravitational wave emission  (HB15). In an effort to understand SMBH evolution in a galactic context, we also simulate the initial inspiral of SMBHs due to dynamical friction, as opposed to simply following the behavior of bound SMBH binaries. Finally, in addition to mapping the effect of initial SMBH binary orientation, we vary the incoming SMBH 
eccentricity and explore models with kinematically-decoupled cores.

The layout of our paper is described as follows; Section 1 outlines our simulation suite, section 2 discusses our results and we conclude with a summary and discussion.

\section{Simulation Setup} \label{mer-sim}

	\subsection{Initial Models \& SMBH Orbital Plane}
	
	To facilitate comparison with HB15, we begin with the same axisymmetric galaxy model, with a flattening parameter, c/a, of 0.8 at the half-mass radius, and an inner density slope, $\gamma=1$. In our model units, the total mass is equal to 1 and the SMBH mass, $M_\bullet$, is  0.005, while the scale radius ($r_0$) is 0.5. The influence radius, defined as the radius of a hypothetical sphere around a SMBH that contains a stellar mass equal to twice that of a SMBH, is 0.08 in our model units and contains 80,000 particles. In our fiducial runs, a second equal-mass SMBH is placed at a distance of 0.5 from the central SMBH with a velocity equal to 63\% of the circular velocity at that radius. Each simulation contains 1M particles. 
	
	\begin{table}{}
		\caption{The simulation suite. The A models explore the effect of misalignment between the binary SMBH and galaxy angular momentum vectors. The B suite is designed to simulate the effect of counter-rotating cores, and the C suite varies the initial SMBH eccentricity in a fully-counterrotating model.\label{tab:parameters}}
        \begin{tabular}{cccc}
        		\hline
		Model & Rotation & $\theta$ & Rotation Radius\\
        		\hline
		$A_{0}$  & Co-rotation & $0^\circ$ & full galaxy\\ 
		$A_{45}$ & Co-rotation & $45^\circ$ & full galaxy\\
		$A_{90}$ & NA & $90^\circ$ & full galaxy\\ 
		$A_{120}$ & Counter & $120^\circ$ & full galaxy\\
		$A_{135}$ & Counter & $135^\circ$ & full galaxy\\ 
		$A_{150}$ & Counter & $150^\circ$ & full galaxy\\ 
		$A_{180}$ or $C_{0.60}$ & Counter & $180^\circ$ & full galaxy\\ 
		$B_0$ & Co-rotation & $0^\circ$ & 0.3\\
		$B_{180}$ & Counter & $180^\circ$ & 0.3\\
		$C_{0.44}$ & Counter & $180^\circ$ & full galaxy\\	
		$C_{0.36}$ & Counter & $180^\circ$ & full galaxy\\	
		$C_{0.19}$ & Counter & $180^\circ$ & full galaxy\\
		$C_{0.00}$ & Counter & $180^\circ$ & full galaxy\\
		\hline
        \end{tabular}

Column 1: Galaxy model. The subscript for the A and B models refer to the initial SMBH orbital plane orientation, $\theta$. The subscript on the C models refer to the initial eccentricity of the secondary SMBH orbit.  Column 2: Rotational sense of SMBH binary initial orbits  with respect to the galaxy. Column 3: Angle ($\theta$) between angular momentum of SMBH binary ($L_{\rm bin}$) with respect to that of the galactic plane ($L_{\rm gal}$). Column 4: Radius of the rotational part of the galaxy in length units.

	\end{table}
	
\begin{figure}
			\centering
			\includegraphics[width=1\linewidth]{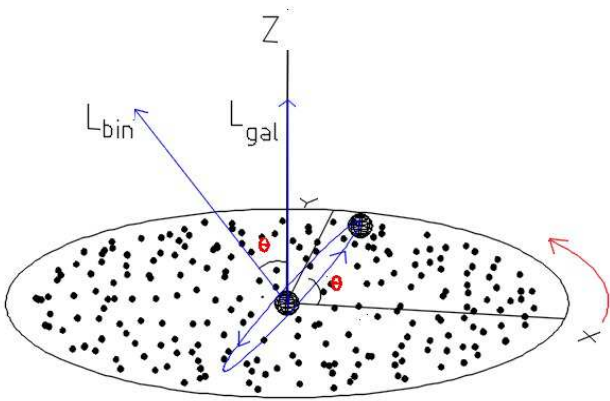}
			\caption{Illustration of the initial angle, $\theta$, between the angular momentum of the SMBH binary ($L_{\rm bin}$) and that of the galaxy ($L_{\rm gal}$). In each case, $L_{gal}$ points along the z-axis and is denoted by a red arrow.  As the simulation starts, the SMBHs (black dots) orbit each other (blue trajectory) within the plane of the galaxy (smaller black dots). The angle, $\theta$, between angular momentum vectors is also naturally the angle between the SMBH binary orbital plane and the rotational plane of galaxy.}
			\label{illustration}
		\end{figure}

	Table \ref{tab:parameters} catalogs the suite of simulations investigated in this study. The A series varies the inclination of the binary orientation with respect to the angular momentum axis of the galaxy, which is uniformly rotating. The B models are designed to explore the effect of rotation within the galaxy core by limiting rotation to only the inner 0.3 radial units. Finally, to investigate the effect of counterrotation on the eccentricity evolution of the binary, the C series varies the initial velocity of the secondary SMBH within a fully counterrotating galaxy. The subscripts on the A and B models indicate the angle between the angular momentum vector of the binary, $L_{\rm bin}$,  and galaxy, $L_{\rm gal}$, while the subscripts on the C models specify the initial eccentricity. Note that with this nomenclature,  the $A_{180}$ and $C_{0.60}$  are identical. Figure \ref{illustration} shows a schematic of the configuration for each run.

	We introduce galaxy rotation by flipping the direction of the velocity components such that $L_{\rm z}$ is positive  for every particle. With the A galaxy model in place, we set the components of the initial velocity of the secondary SMBH such that the total $L_{\rm bin}$ has the desired $\theta$. For instance, in the case of $A_{45}$, the velocity was equally distributed between its y and z components. To produce the B galaxy model, we set $L_{\rm z}$ to be positive only for the particles within a radius of 0.3 from the galaxy center. Density profiles of the A and B models at T= 0, 2 and 6 time units confirm its stability in terms of structure; kinematically, the B model experienced some minor evolution at the interface between the outer galaxy and the kinematically distinct core, as expected, due to orbital mixing between the two regions. Within 0.2 radial units, however, the initial velocity profile is stable.


\subsection{Numerical Code and Hardware}

In this work, we use the fully parallelized direct $N$-body code $\varphi$GPU ~\citep{BNZ2011}, which uses only CUDA and MPI to optimize gravitational force calculations on GPU hardware, achieving well over a factor of two speedup compared to multi-core hyperthreaded CPU architectures.

The code is a direct $N$-body simulation package, with a 
Hermite integration scheme from 4-8th order and individual block time steps. For this study, we determined that the 4$^{\rm th}$ order integration scheme served our purposes. In principle, a direct $N$-body code evaluates all pairwise forces between the gravitating particles, but $\varphi$GPU can introduce softening to avoid gravitational force divergence. The softening value for particles was 
taken to be $10^{-4}$ while it was zero for SMBHs. We refer more interested readers to a general discussion about direct $N$-body codes and their implementation in \cite{spurzem2011a,spurzem2011b}.
 
More details on the public version of $\varphi$GPU are presented in \cite{BNZ2011, SBZ2012, Berczik2013}. Our non-public version of the code is well-tested and has been used in several published multi-million particle simulations~\citep{KBB2012, Khan+16}.

We ran our experiments using 48 GPUs in the Advanced Computing Center for Research and Education $($ACCRE$)$ at Vanderbilt University.


\section{SMBH Coalescence in Rotating Galaxy Models} \label{BBH-evo}

	We now present the results of our A,B, and C simulation suites (mentioned in Table. \ref{tab:parameters}), focusing in particular on the behavior of the eccentricity, inverse semi-major axis, energy and angular momentum loss, and the SMBH center of mass. We synthesize these findings to predict the SMBH binary coalescence timescale. 
    
    \subsection{Eccentricity evolution}
    
\begin{figure}
			\centering
			\includegraphics[width=1\linewidth, height=2.25in]{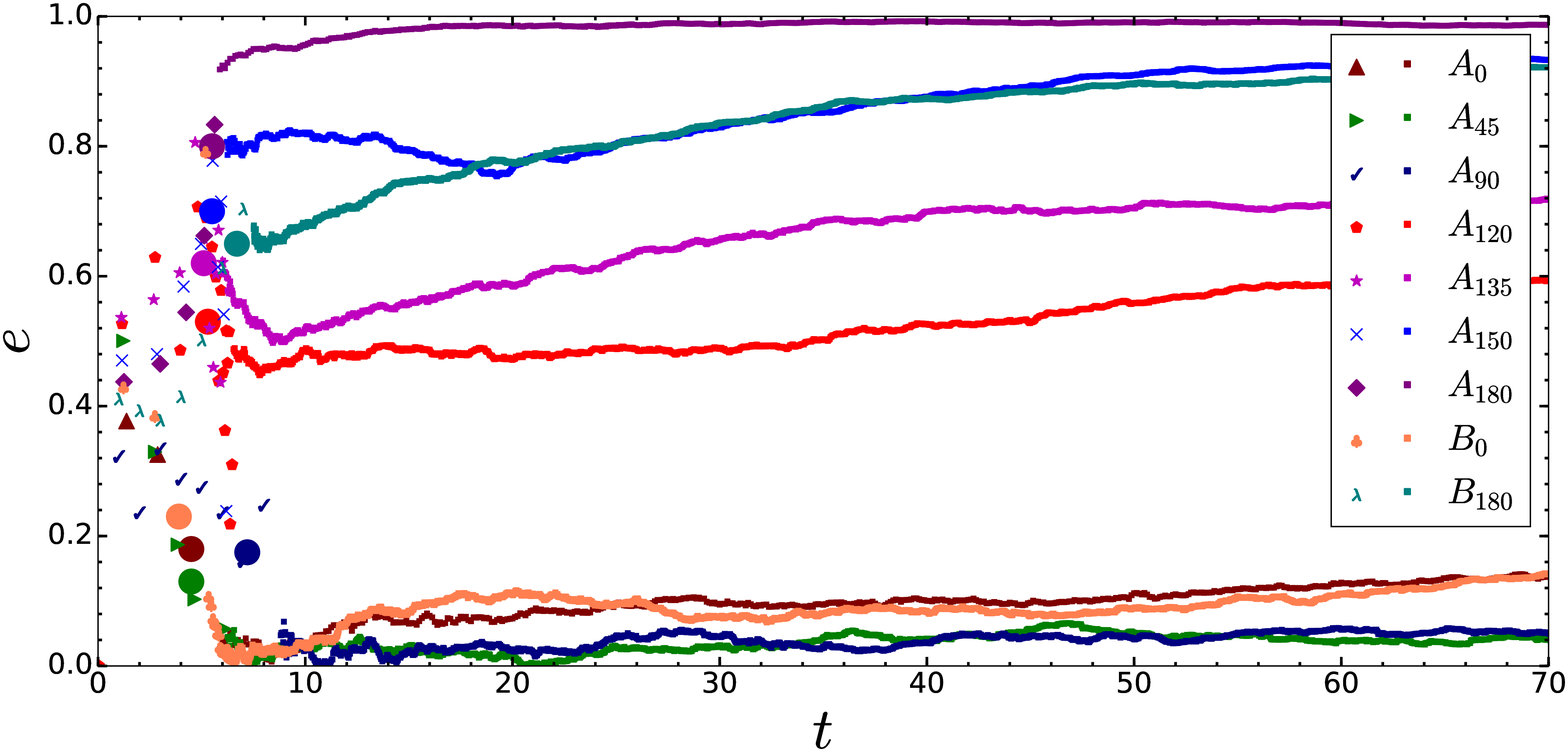}
			\caption{Evolution of the eccentricity for A and B models. Notice the decrease in eccentricity from $A_{180}$ to $A_{90}$. For initially unbound SMBHs , we calculate the eccentricity using equation \ref{eq:relative} and represent it with various symbols on the left side in the legend. As the binary is formed (indicated by filled circles), equation \ref{eq:eccn-kepler} is employed for subsequent hardening phase, denoted with boxes ($\blacksquare$).}
			\label{fig:eccentricity}
		\end{figure}

    First, we feature the evolution of the SMBH binary eccentricity for the A and B models in figure \ref{fig:eccentricity}. During the initial phase both SMBHs are just moving about one another in an unbound orbit, the eccentricity of their trajectory can still be computed using:
	\begin{equation}
		e=\frac{r_a - r_p}{r_a+r_p}
		\label{eq:relative}
	\end{equation}
\noindent where $r_a$ and $r_p$ are the separation distances between the SMBHs at apoapsis and periapsis, respectively. As the two SMBHs form a bound system marked by filled circles in figure \ref{fig:eccentricity}, we use the Keplerian definition of eccentricity:
		\begin{equation}
		e=\sqrt{1+\frac{2\epsilon h^2}{\mu^2}}
		\label{eq:eccn-kepler}
		\end{equation}
\noindent where $\epsilon$ is the specific orbital energy, h is the specific relative angular momentum and $\mu$ is the standard gravitational parameter.
	
Broadly speaking, the final eccentricity of the binary orbit is low for corotating models and high for counterrotating ones~\citep{Dotti+07,Sesana+11, Madigan+12}, though we find that the eccentricity evolution is non-monotonic and describe it more below.

	The eccentricity is persistently high for all cases of counterrotation, regardless of the inclination of the orbital plane. Indeed, the final eccentricity for these counterrotating runs is nicely correlated with $cos \theta$ as the projection of the opposing angular momentum decreases. As the secondary black hole infall toward the center, the eccentricity rises because interactions with retrograde orbits are so efficient at extracting angular momentum. Once the binary forms, though, there is a brief episode of circularization for the $A_{150}$, $A_{135}$ and $A_{120}$ runs that coincides with a reorientation of the binary orbital plane to a more prograde configuration (see the temporary dip in eccentricity shortly after binary formation, marked by the filled circles in Figure 2). This momentary gain of binary angular momentum is lost again as the binary hardens and ejects prograde orbits.
    
    On the other hand, the final eccentricity remains pretty close to 0 for initially corotating models, namely $A_{0}$, $A_{45}$ and $A_{90}$. As in the counterrotating runs, we found that the binary orbital angular momentum vector reorients to better align with that of the galaxy so that it has a larger corotating component.  In these corotating cases, the eccentricity of the secondary SMBH orbit initially decreases in the classic behavior of dynamical friction to circularize an orbit\citep[e.g.][]{Dotti+07,Antonini+12,Petts+15}; in our case, the high central density near the cusp makes dynamical friction especially strong near pericenter, despite the high velocity.
    

	A look at the centrally-rotating $B_0$ and $B_{180}$ models reveals very similar behavior as the A simulation suite: the binary orbit is nearly circular for the corotating case ($B_0$), and is highly radial when the galaxy model is counter-rotating ($B_{180}$). This shows that the kinematics at the galaxy center, rather the outer suburbs, have the most influential effect on the SMBH binary orbit.  Although we did not explore even smaller decoupled cores, we speculate that this strong eccentricity evolution will continue until the enclosed galaxy mass is a few times the mass of the SMBH binary.

\begin{figure}
			\centering
			\includegraphics[width=1\linewidth, height=2.25in]{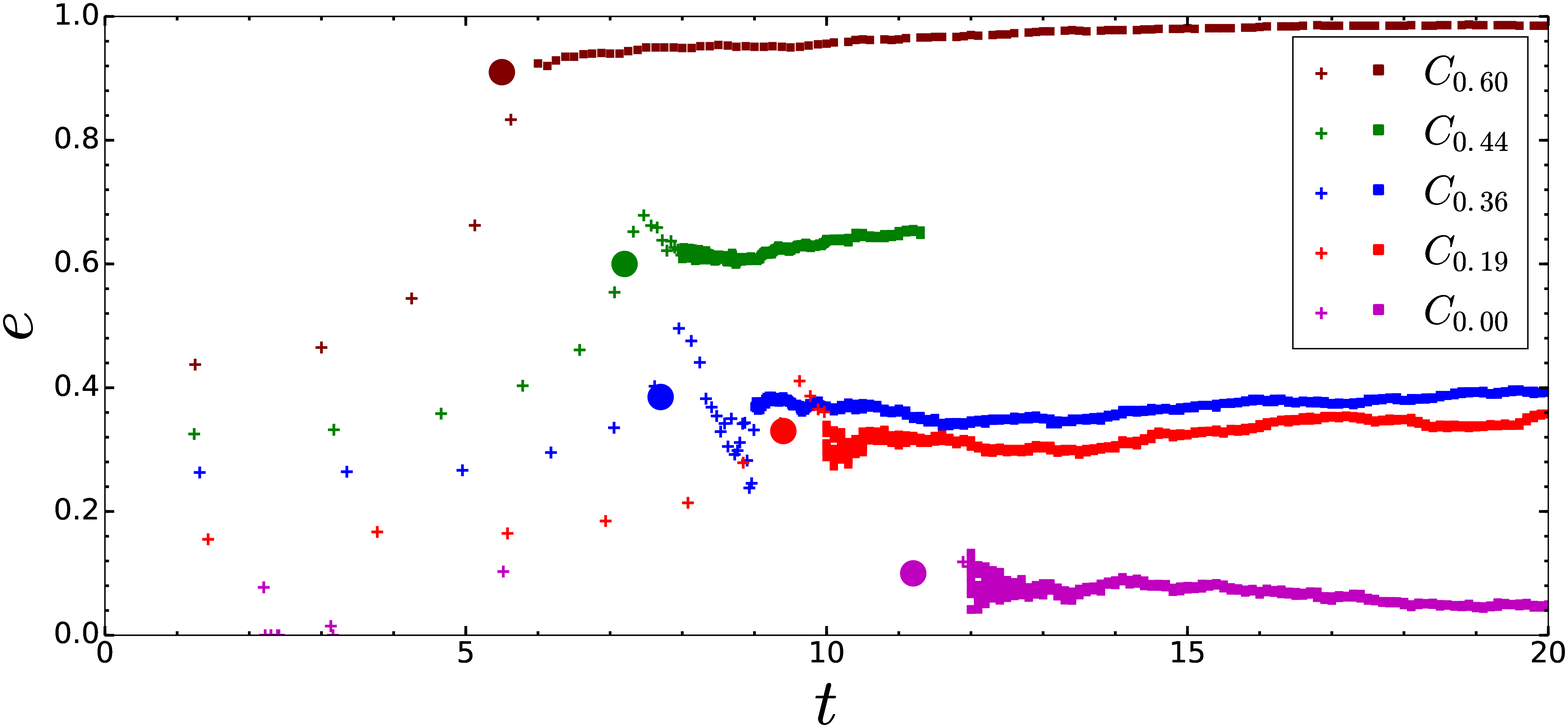}
			\caption{Eccentricity evolution for C models with varying initial eccentricity. The cross (+) symbol has been used to denote the initial phase, where eccentricity is determined by  equation \ref{eq:relative}, while the box ($\blacksquare$) symbol has been used for the binary. Filled circles note the time of binary formation.
				}
			\label{fig:eccen-ini_vel}
		\end{figure}

	Turning now to the C simulation suite, which tracks the SMBH binary evolution within counterrotating models as a function of initial eccentricity, we find that the eccentricity decreases initially from its starting value and then increases rapidly just before a Keplarian SMBH binary forms.
This relatively abrupt phase of eccentricity evolution changes to more stable evolution during 3-body scattering phase. Table \ref{tab:ini-vel} describes the initial velocity and eccentricity of the secondary SMBH orbit. In figure \ref{fig:eccen-ini_vel}, recall that eccentricity is initially calculated using eq. \ref{eq:relative},  while for remaining time, we used eq. \ref{eq:eccn-kepler}. It is clear that the initially circular model ($C_{0.00}$) remains circular in later phases, but the nearly-circular ($C_{0.19}$ and $C_{0.36}$) binaries enter the hardening phase with a slightly higher than initial eccentricity.  Likewise, initially eccentric orbits of $C_{0.44}$ and $C_{0.60}$, become more eccentric, with the final eccentricity upon entering the hardening phase of 0.65 and 0.99, respectively. 

 Our results support the behavior of the in-plane setups -- $A_{0}$ and $A_{180}$ -- of HB15 and are consistent with prior work on eccentricity evolution in a variety of stellar systems with different mass ratios, kinematic and structural setups than ours~\citep[e.g.][]{Sesana+11, Iwasawa+11, Madigan+12, Wang+14}. These works conclude that preferential ejection of prograde stars can help to pump SMBH binary eccentricity up to unity in the continuum limit, though secular processes in rotating systems may act to amplify eccentricity growth or decay~\citep{Madigan+12}.  Our work is consistent with the findings of \citet{Iwasawa+11}, who noted in their Figure 4 that the initial reservr of prograde orbits are quickly scattered; thereafter, retrograde orbits may be scattered into prograde orbits by the SMBH binary, after which they may be ejected from the system. However, we may also interpret our eccentricity growth in terms of {\em secular dynamical anti-friction}, a long-term net torque arising from  the ensemble of retrograde orbits~\citep{Madigan+12}. As in \citet{Sesana+11}, we find that the eccentricity rise is more steep for binaries with initial eccentricity greater than 0.5. However we notice that binaries with an initial $e<0.5$ do not reach $e\sim 1$, and circular binaries remain close to circular. Most of the eccentricity evolution occurs within a few orbital times of the SMBH binary formation. Our rapid eccentricity evolution seems to be in contrast with \citet{Sesana+11} where the eccentricity approached unity in about 100 orbital times,  however the faster evolution in our study may owe to a longer initial secondary orbital timescale, which would allow the secondary to interact with a larger mass reservoir early on.  We caution that since the eccentricity evolution depends so strongly on the angular momentum admixture of individual strong encounters with particles in the stellar model, as well as any secular perturbation on the stellar background excited by the secondary black hole motion, higher resolution simulations may be required in the future to ensure numerical convergence.

\begin{table}{}
    \centering
		\caption{Initial velocities and eccentricity evolution of the C suite.\label{tab:ini-vel}}
		\begin{tabular}{ccccc}
		\hline
		Model & $x=\frac{v_{sec,i}}{v_{cir}}$  & $e_{i}$ & $e_{bin}$ & $e_{hard}$\\
		\hline
	$C_{0.60}$   & 0.63  & 0.60 & 0.9 & 0.99 \\ 
	$C_{0.44}$   & 0.75  & 0.44 & 0.55 & 0.65 \\ 
	$C_{0.36}$  &  0.80  & 0.36 & 0.35 & 0.35\\
	$C_{0.19}$  & 0.90  & 0.19 & 0.3 & 0.3\\ 
	$C_{0.00}$ &   1.0   & 0.0  & 0.1 & 0.05\\
		\hline
		\end{tabular}

Column 1: Galaxy model.  Column 2: Fraction of circular velocity, given to secondary SMBH initially, to obtain desired eccentricity. Column 3: Initial eccentricity of the (unbound) secondary SMBH orbit. Column 4: Eccentricity upon time of bound SMBH binary formation. Column 5: Eccentricity of the hard SMBH binary.
\end{table}

\subsection{Angular momentum loss}

	\begin{figure}
		\centering
		
		\includegraphics[width=1\linewidth]{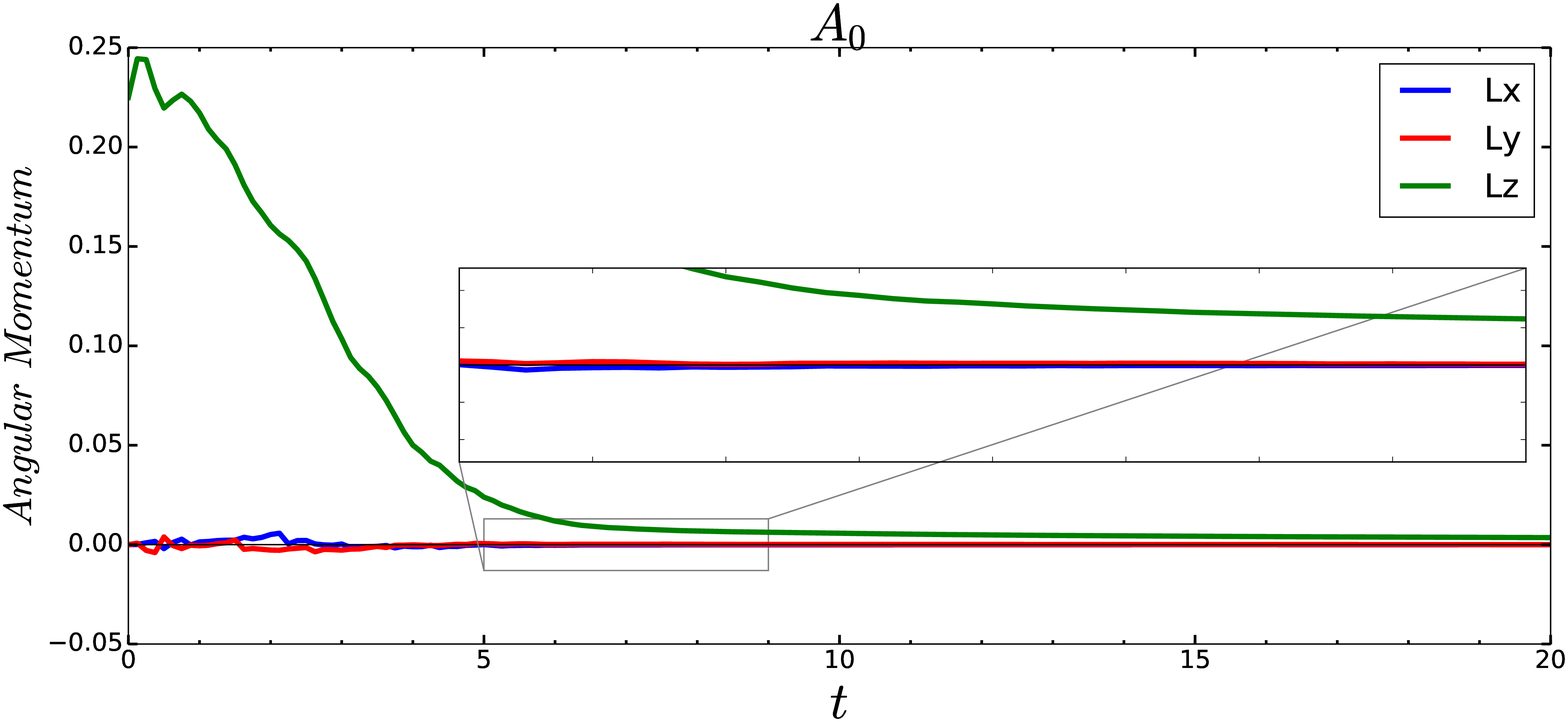}
		\includegraphics[width=1\linewidth]{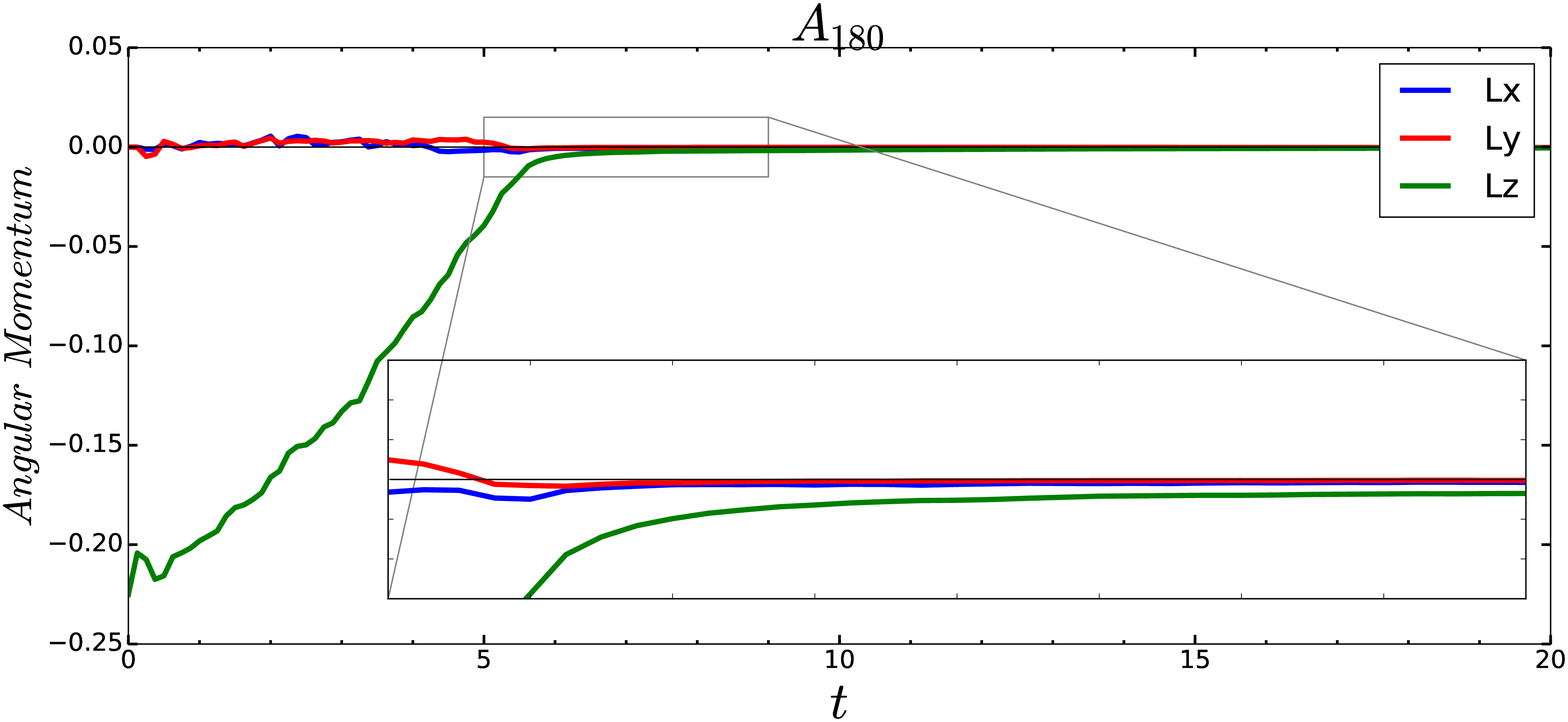}
		\caption{Angular momentum evolution of the SMBH binary for $A_0$ and $A_{180}$ models of HB15.  $A_0$ is at the top and $A_{180}$ is at the bottom.}
		
		\label{fig:angularA0andA180}
	\end{figure}
	
	\begin{figure}
		\centering
		\includegraphics[width=1\linewidth]{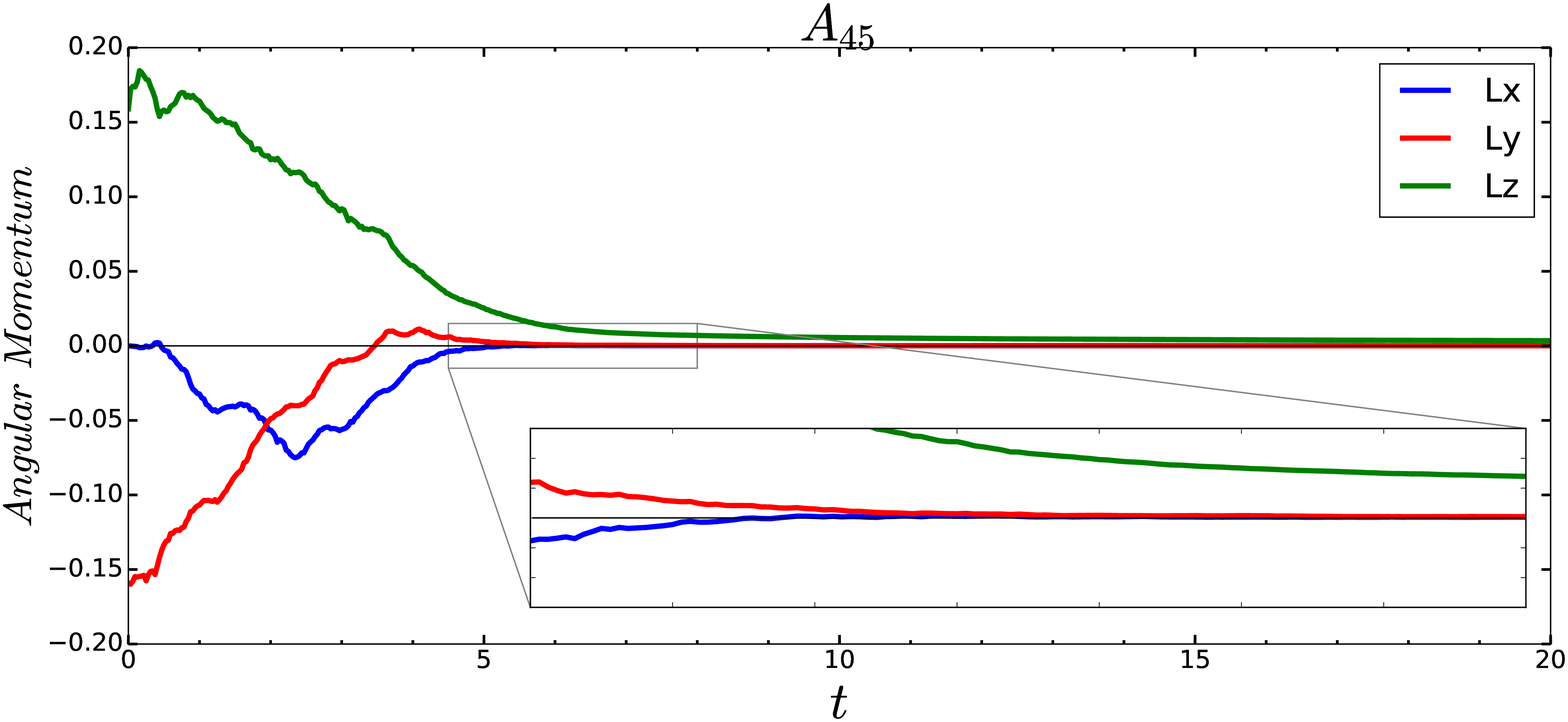}
		\includegraphics[width=1\linewidth]{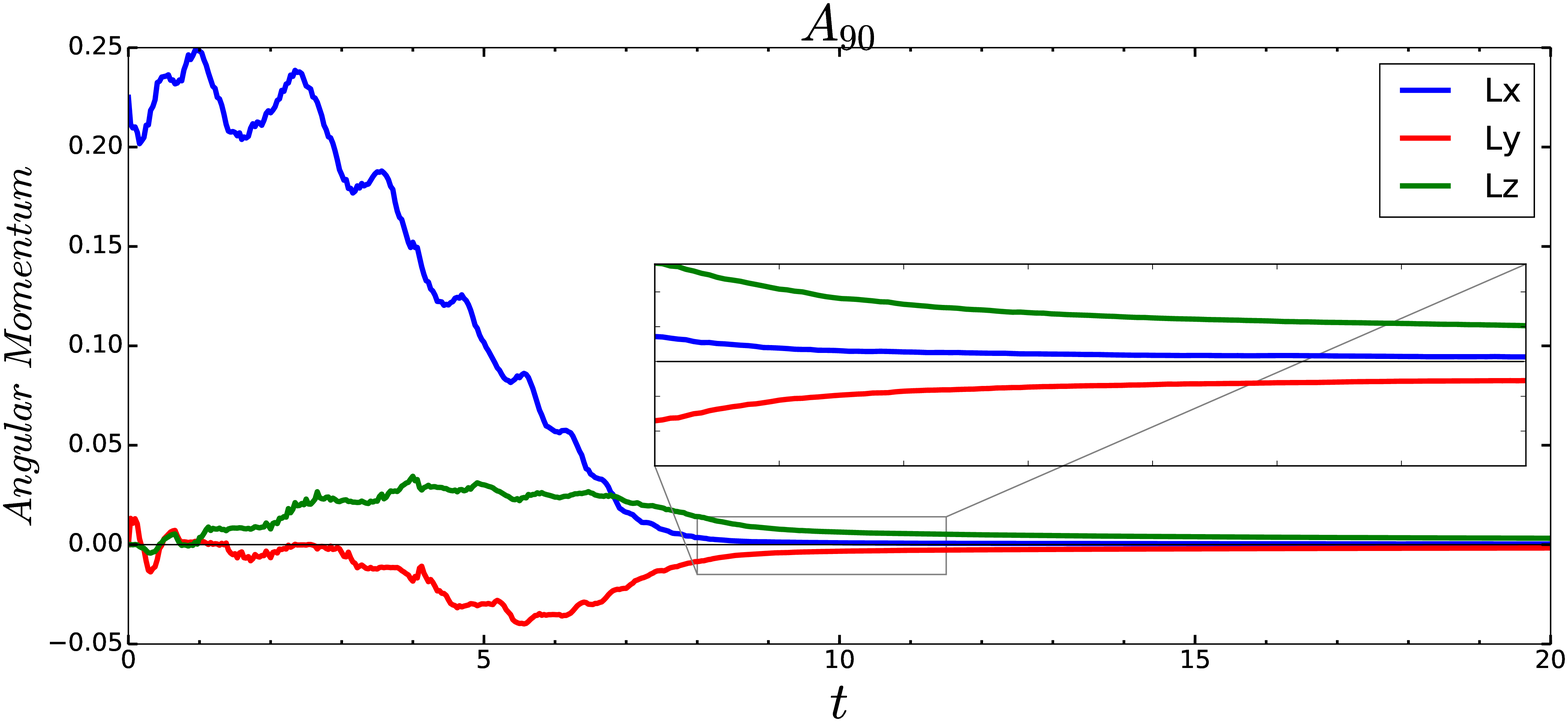} 	
		\includegraphics[width=1\linewidth]{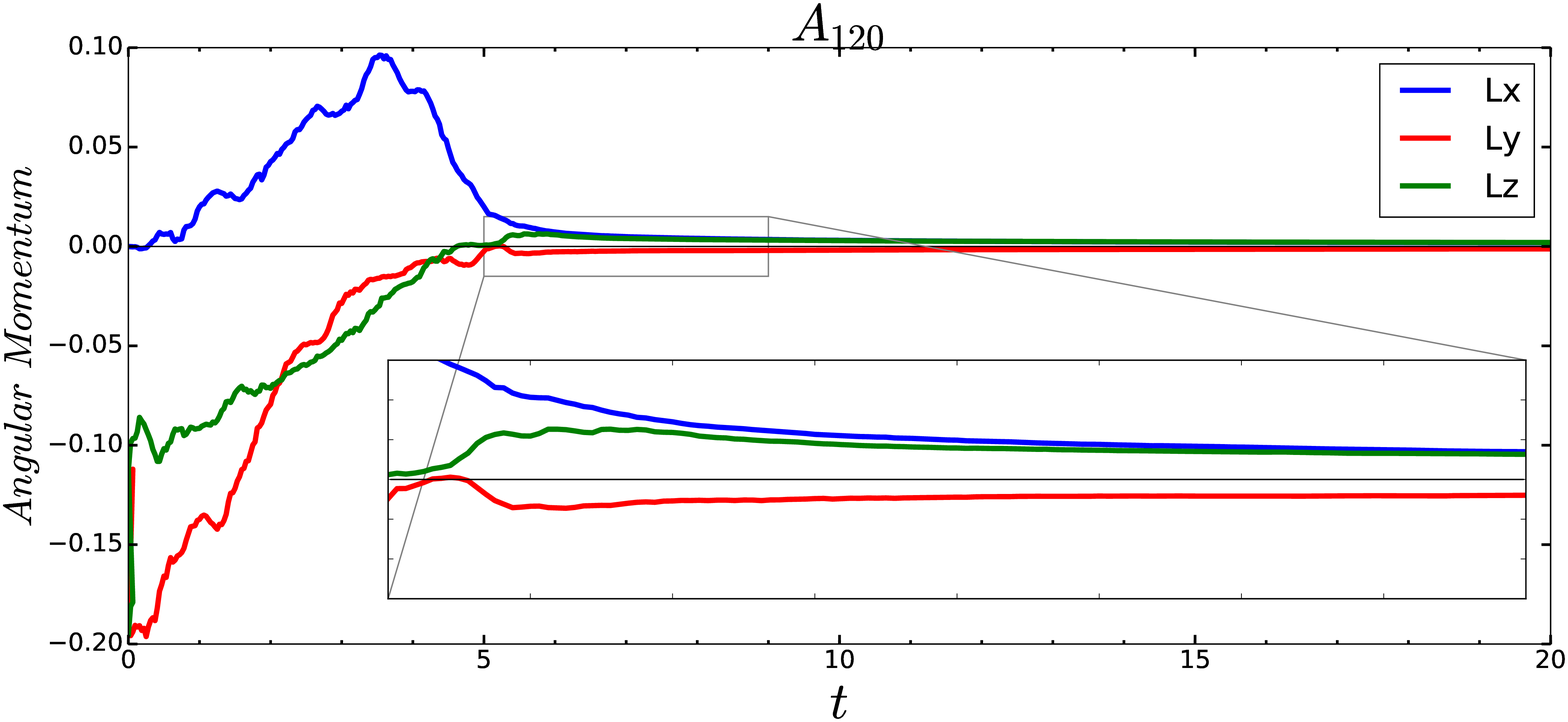}
		\includegraphics[width=1\linewidth]{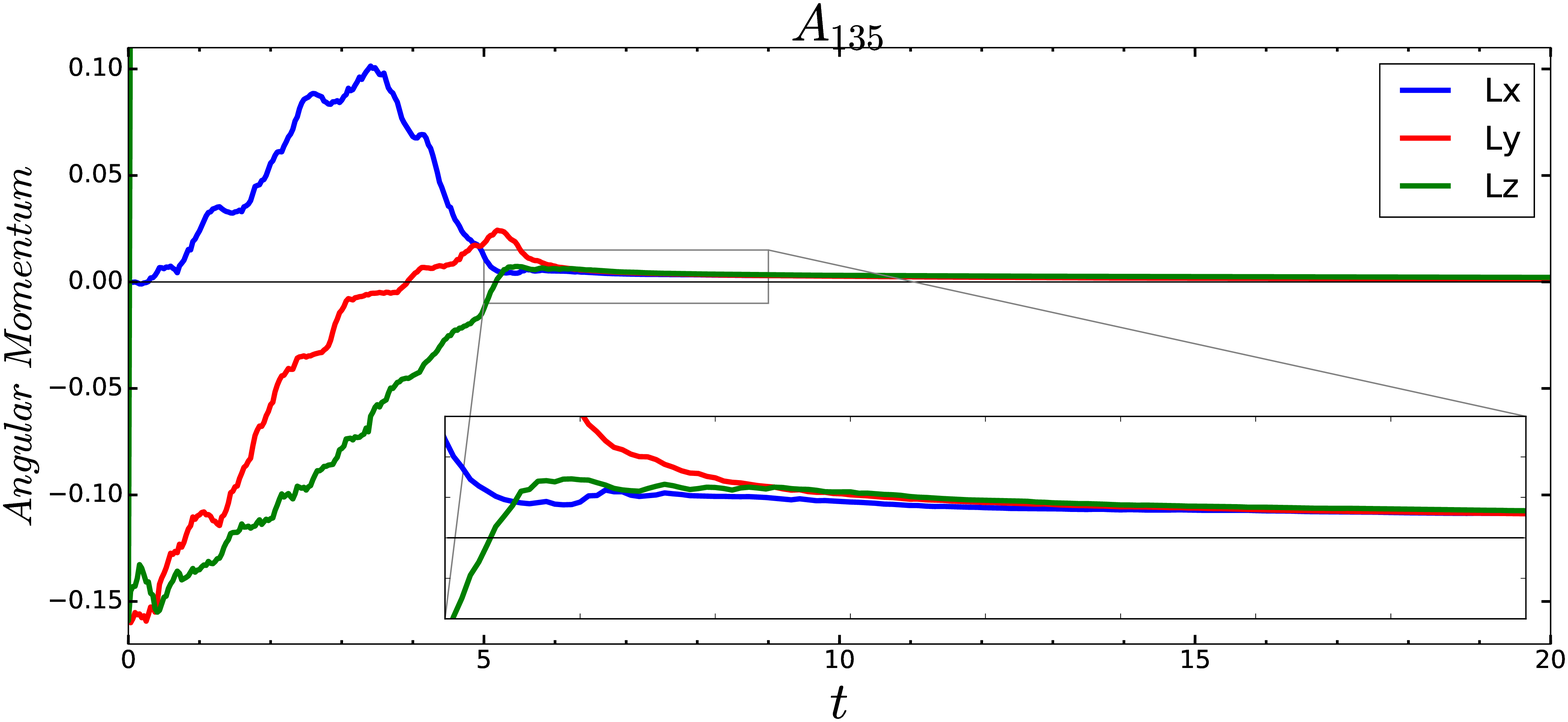}
		\includegraphics[width=1\linewidth]{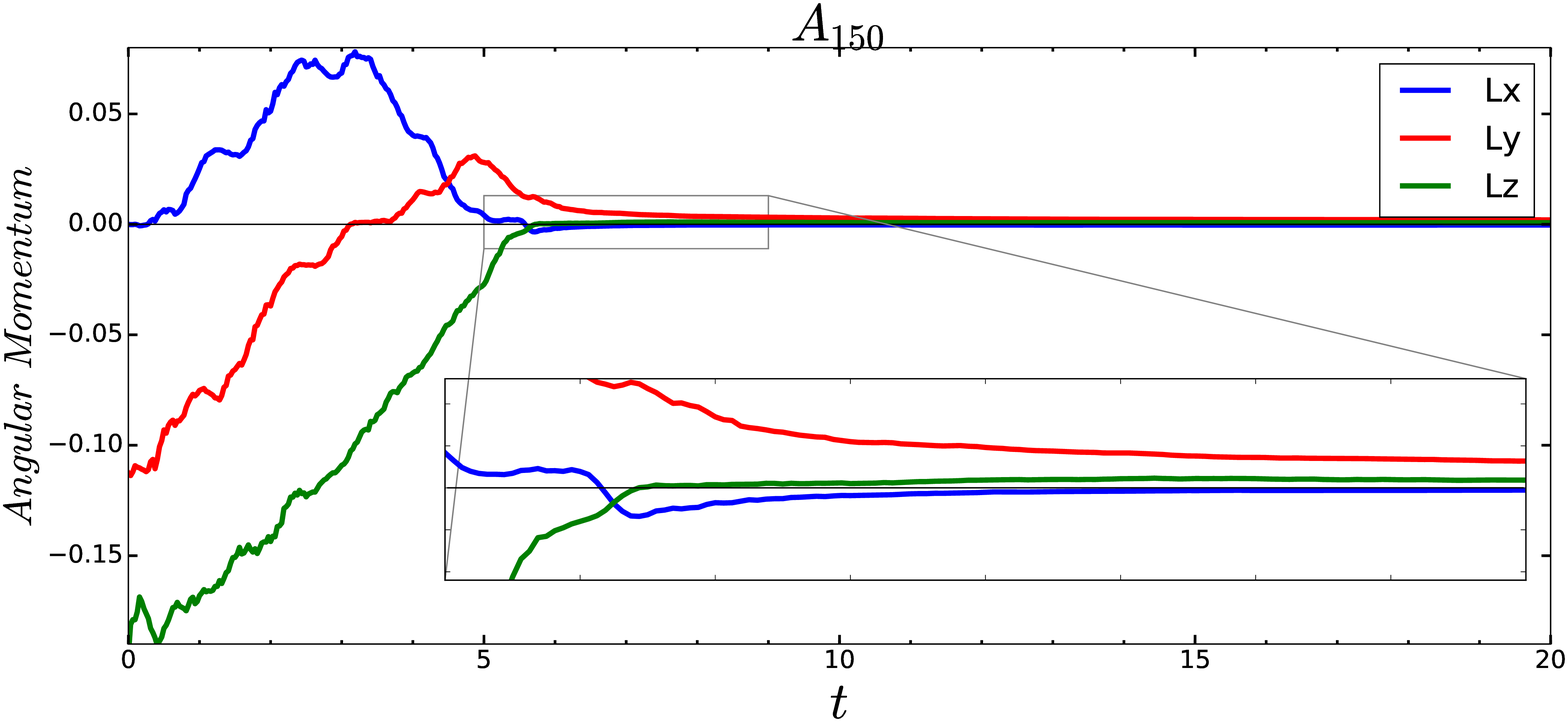}
		\caption{Angular momentum evolution of the SMBH binary with for $A_{45}$ to $A_{150}$  models. From top to bottom: $A_{45}$, $A_{90}$, $A_{120}$, $A_{135}$ and $A_{150}$.}
		\label{fig:angularA45toA150}
	\end{figure}
	
	\begin{figure}
		\centering
		
		\includegraphics[width=1\linewidth]{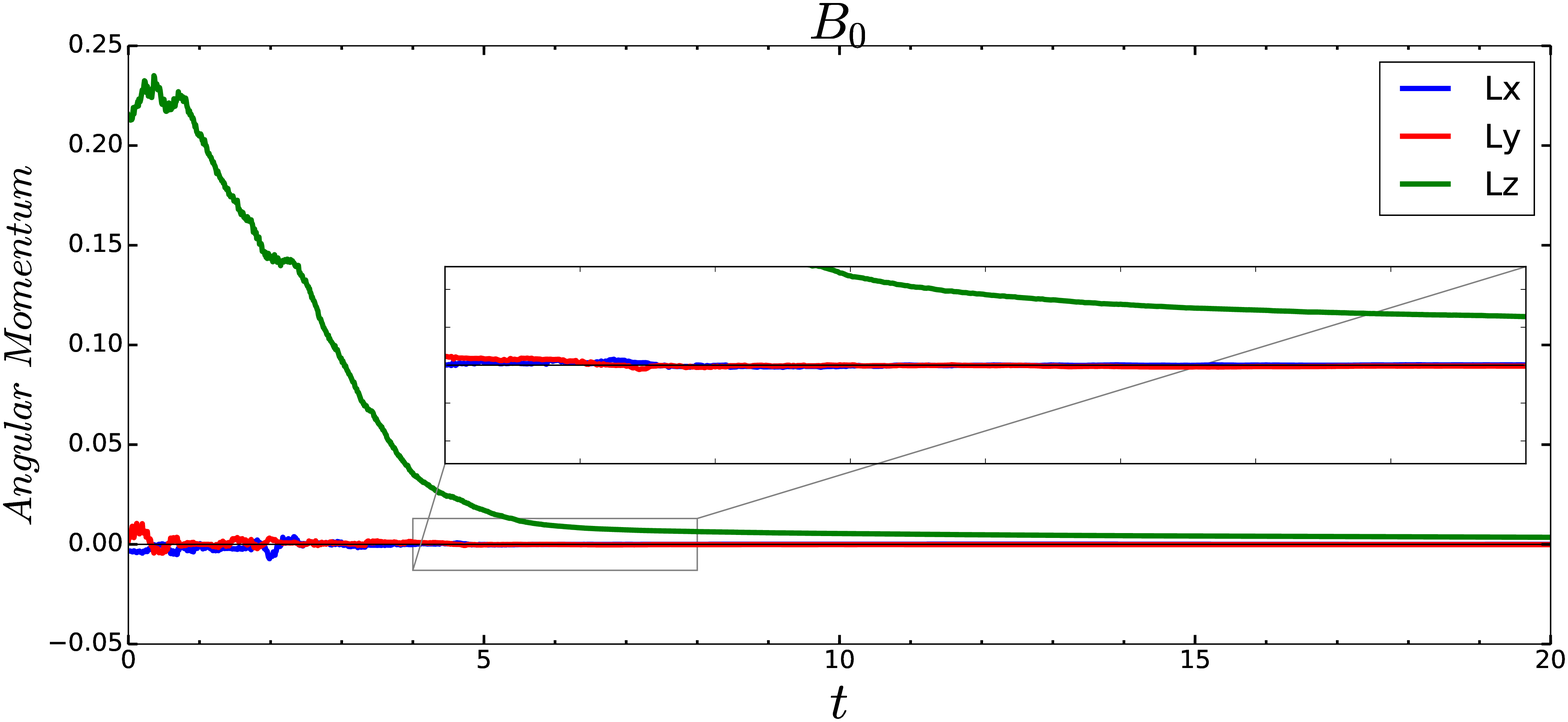}
		\includegraphics[width=1\linewidth]{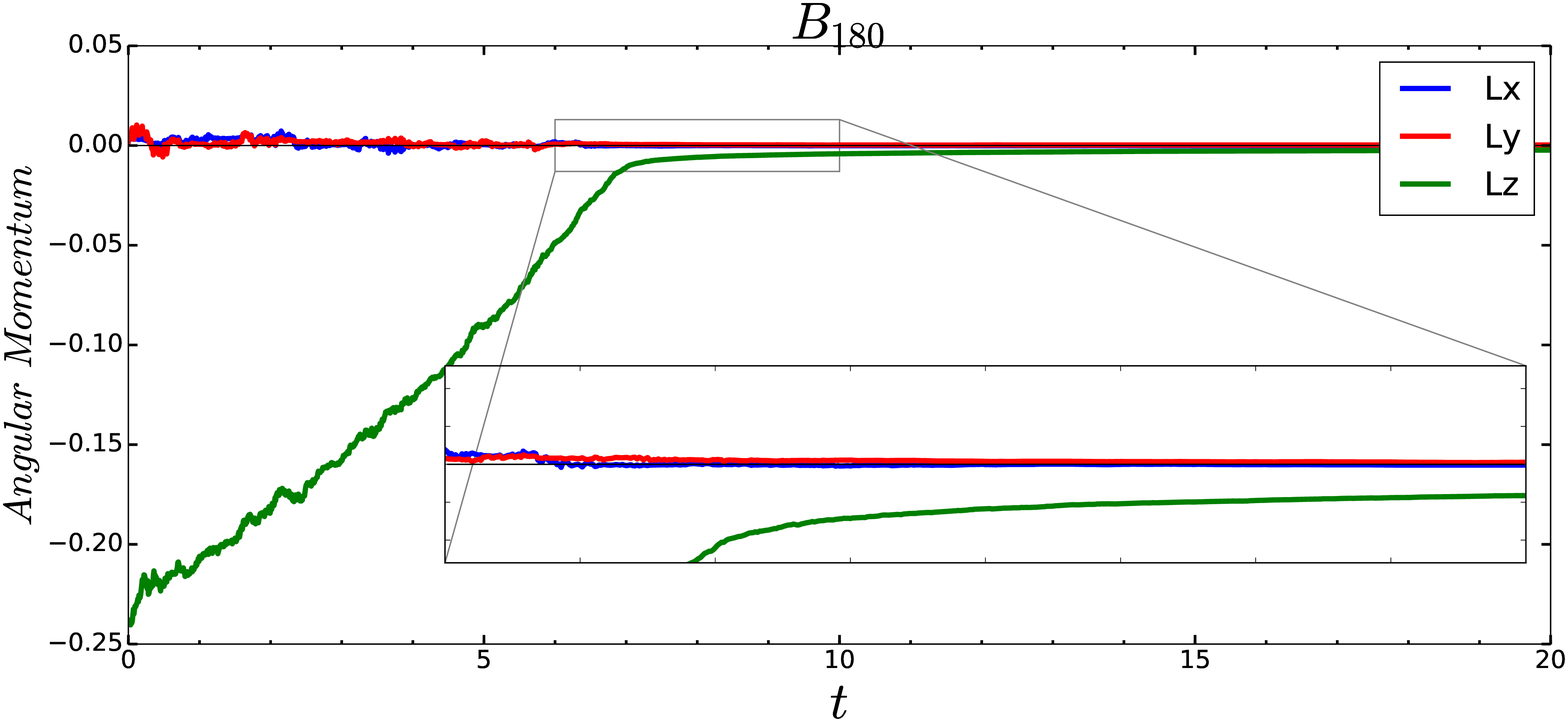}
		\caption{Angular momentum evolution of the SMBH binary for partially rotating B models. $B_{0}$  is at the top and $B_{180}$ at the bottom.} 	
		\label{fig:angularB0andB180}
	\end{figure}

In support of our eccentricity results, we discuss the evolution of the binary angular momenta. Figure \ref{fig:angularA0andA180} reviews the $A_0$ and $A_{180}$ cases of HB15 while figure \ref{fig:angularA45toA150} and figure \ref{fig:angularB0andB180} present the remaining A and B models.
	
\begin{figure}
		\centering
		\includegraphics[width=1\linewidth, height=2.25in]{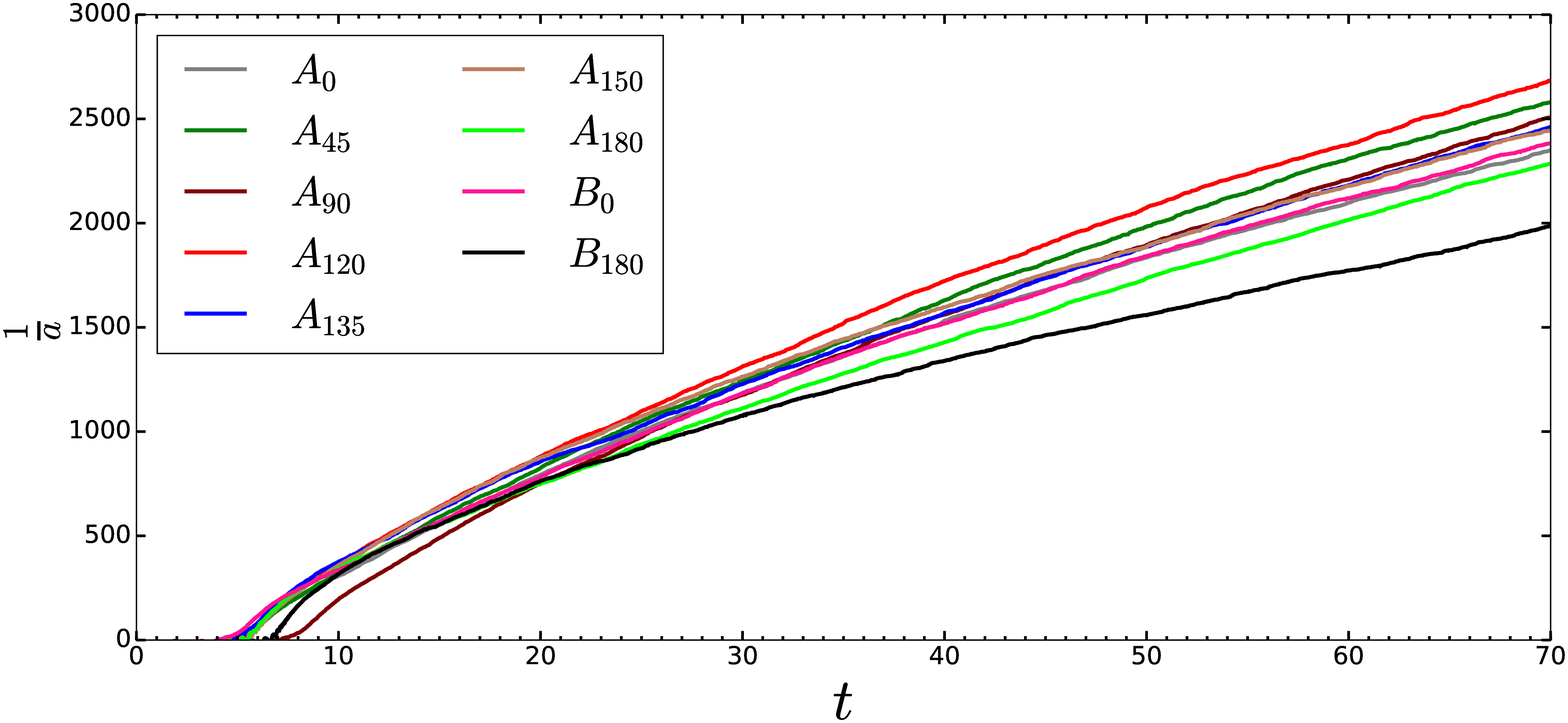}
		\caption{Behaviour of the semimajor axis for the A and B runs. Overlaid are the planar cases (gray and light green) from HB15 for comparison.}
		\label{fig:energy-loss}
	\end{figure}

	The earlier observation that highly eccentric orbits exist in counterrotating cases is substantiated by these figures. Here, we notice that angular momentum quickly reduces for counterrotating cases, whereas figure \ref{fig:energy-loss} points that energy loss rate is roughly equal and comparatively slow. Hence, the rise of eccentricities of counterrotating binaries should indeed be attributed to rapid angular momentum loss. 
	
	Interestingly, in the corotating $A_{45}$ model, the orbital plane flips from $45^\circ$ to $0^\circ$, aligning with the plane of the galaxy. In the counterrotating $A_{135}$, the orbital plane flips over, so that it is corotating at an inclination of $45^\circ$ with respect to the galaxy plane. This shift to corotation is repeated by our remaining counterrotating cases, namely $A_{120}$ and $A_{150}$.  In case of $A_{90}$, the binary orbital plane flips from being perpendicular to the galaxy plane to roughly a  $45^\circ$ inclination. Given such a strong response to align the binary plane with the galaxy, we extrapolate that most galaxies feature corotating SMBH binaries, unless the binary is so radial or overmassive that it enters the gravitational wave regime without significant interaction with the stellar background ~\citep{Khan+15}. A similar alignment between the SMBH binary and galaxy rotation plane was first witnessed by \citet{Gualandris+2012} for high mass ratio binaries, and was argued to have been caused by the same strong prograde encounters that also excite the eccentricity evolution. Here we generalize this finding, witnessing total re-alignment for corotating binaries and partial re-alignment for initially counterrotating binaries before they enter the gravitational wave regime.

	As before, the partially rotating $B_0$ and $B_{180}$ model mirror the A series analogs, further amplifying the concept that the central rotation has the most significant role on the SMBH binary parameters.

    \subsection{Shrinking the binary orbit}
	
	The result with perhaps the greatest import is the hardening timescale, typically represented by the time evolution of the inverse semimajor axis of the SMBH binary (1/a).  Figure \ref{fig:energy-loss} presents this for both the A and B model suites~\footnote{Note that since we ran the C models only to the formation of a hard SMBH binary to explore the eccentricity evolution, we exclude the C suite from analysis of the hardening timescale}. Since the energy of the SMBH binary varies inversely with semi-major axis, these plots can also be seen as tracking of energy extraction. We see the highest energy loss (hardening rate) for mildly corotating model $A_{45}$, whereas the lowest rate is for the counterrotating binary, $B_{180}$. Notice that there is no distinct pattern of energy loss that could be associated with angular variation or precession.

	We observed higher hardening rates for the corotating configurations $A_{45}$ and $B_{0}$ compared to their counterrotating binaries partners, $A_{135}$ and $B_{180}$. Similar behavior was reported in HB15 for corotating $A_{0}$ vs counterrotating $A_{180}$ models. The higher energy loss in corotating models could be due to the motion of the SMBH binary about the galactic center. By settling into a 1:1 resonance with the galactic orbital period, the SMBH binary has more stars available to scatter and remain in this stellar reservoir longer.  
Finally, the similarity of hardening timescales for  $A_0$ and $B_0$, as well as for $A_{180}$ and $B_{180}$ supports our earlier assertion that the innermost stars are responsible for SMBH binary evolution.

\subsection{Binary orbital trajectory}

	\begin{figure}
		\centering
		\includegraphics[width=1\linewidth, height = 1.7in]{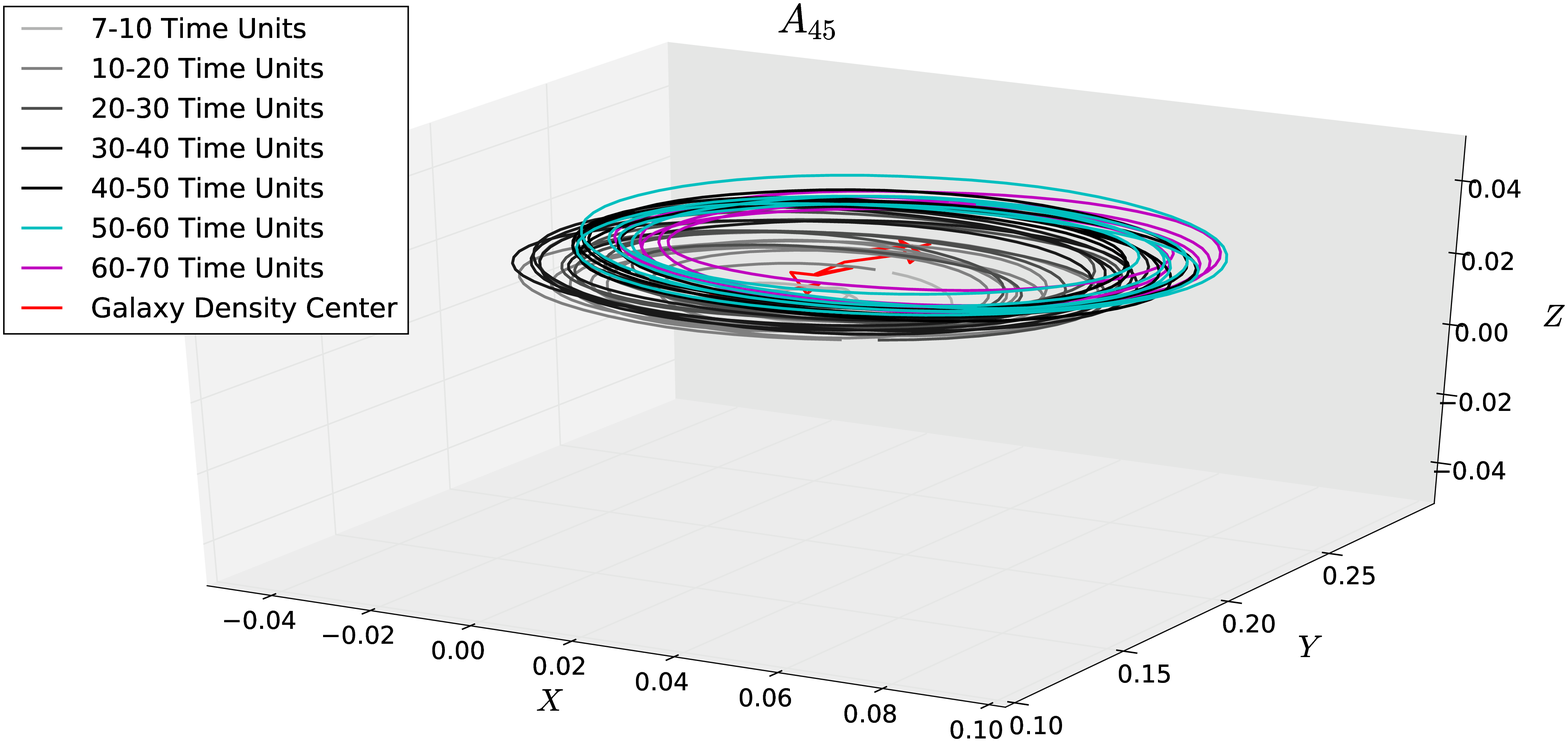}
		\includegraphics[width=1\linewidth, height = 1.7in]{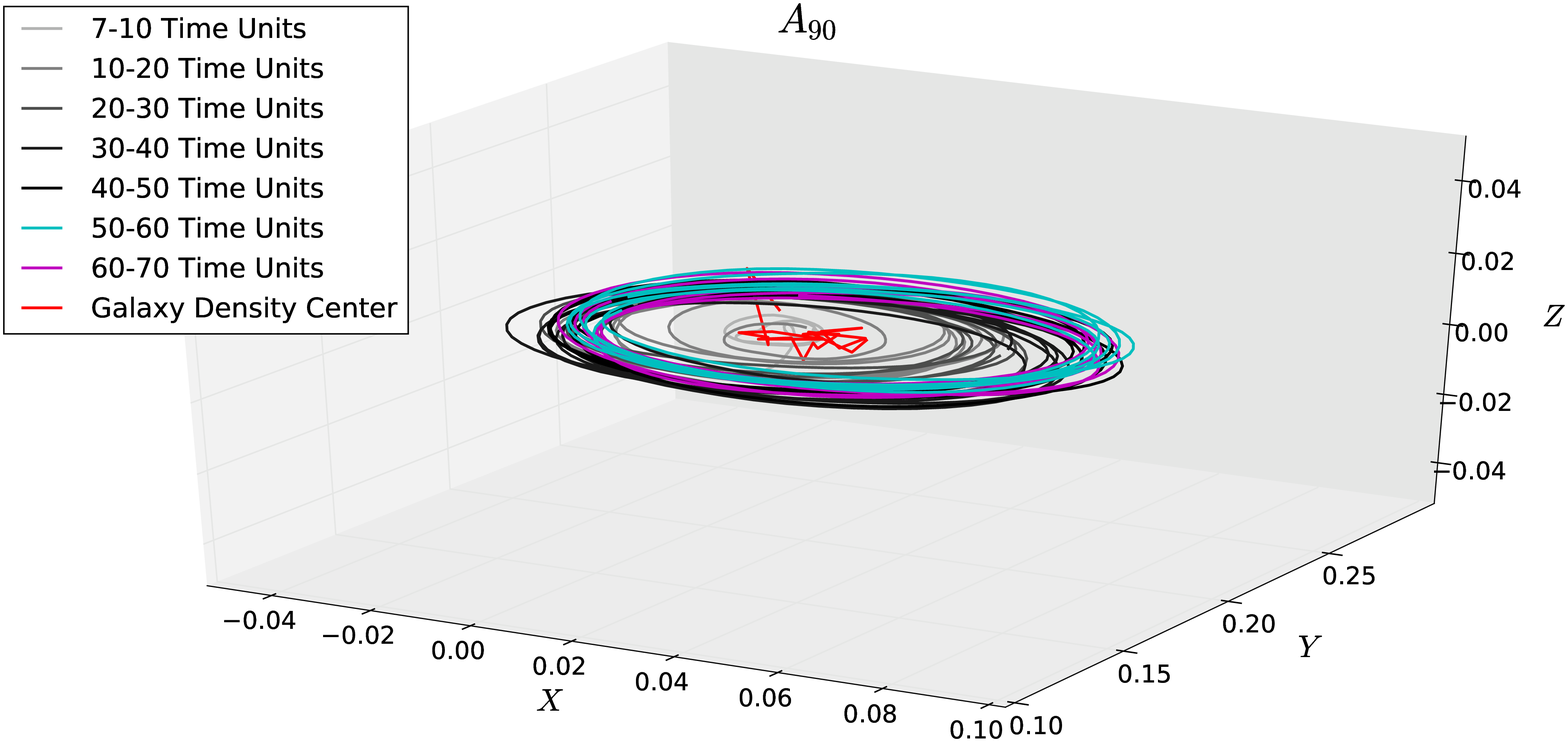}
		\includegraphics[width=1\linewidth, height = 1.7in]{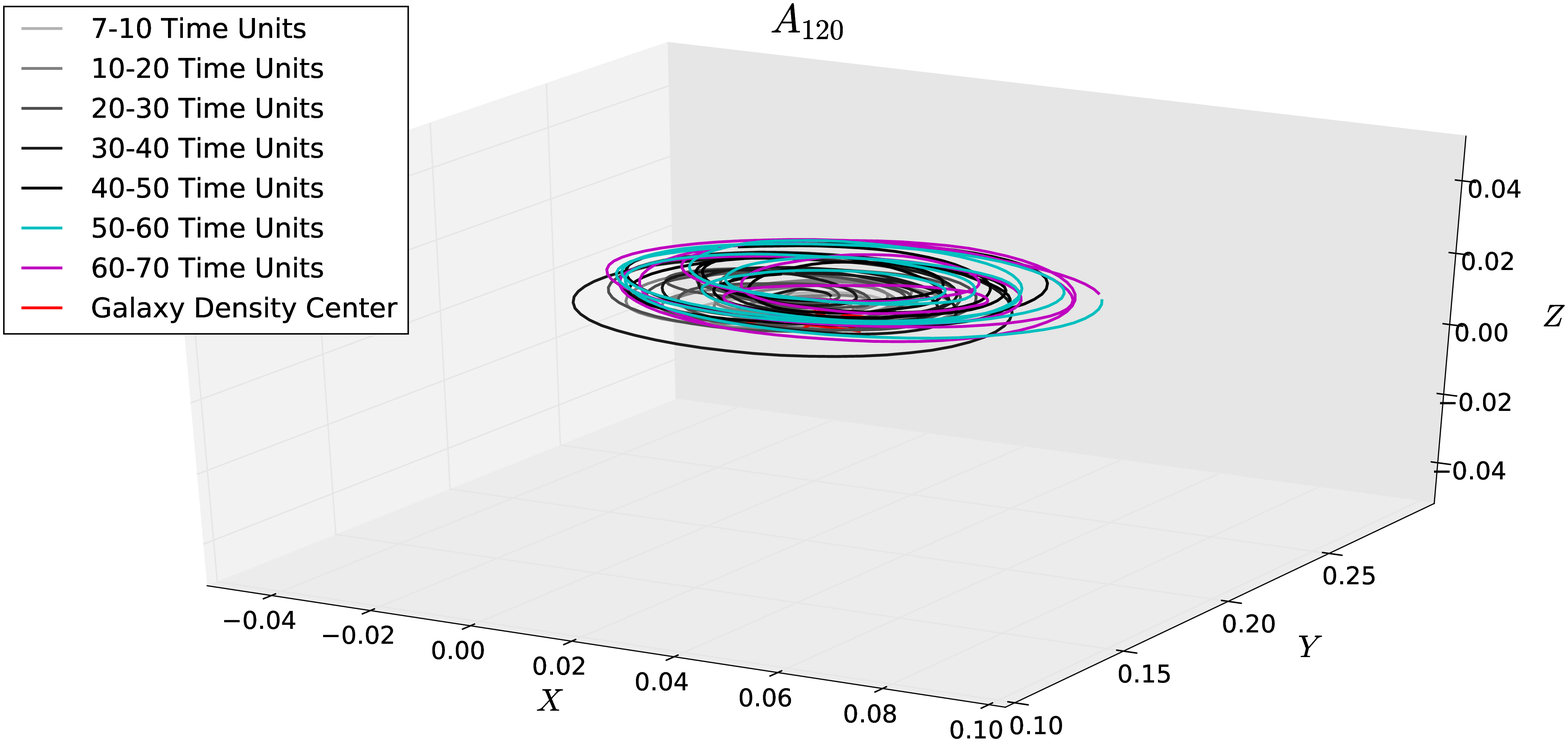}	
		\includegraphics[width=1\linewidth, height = 1.7in]{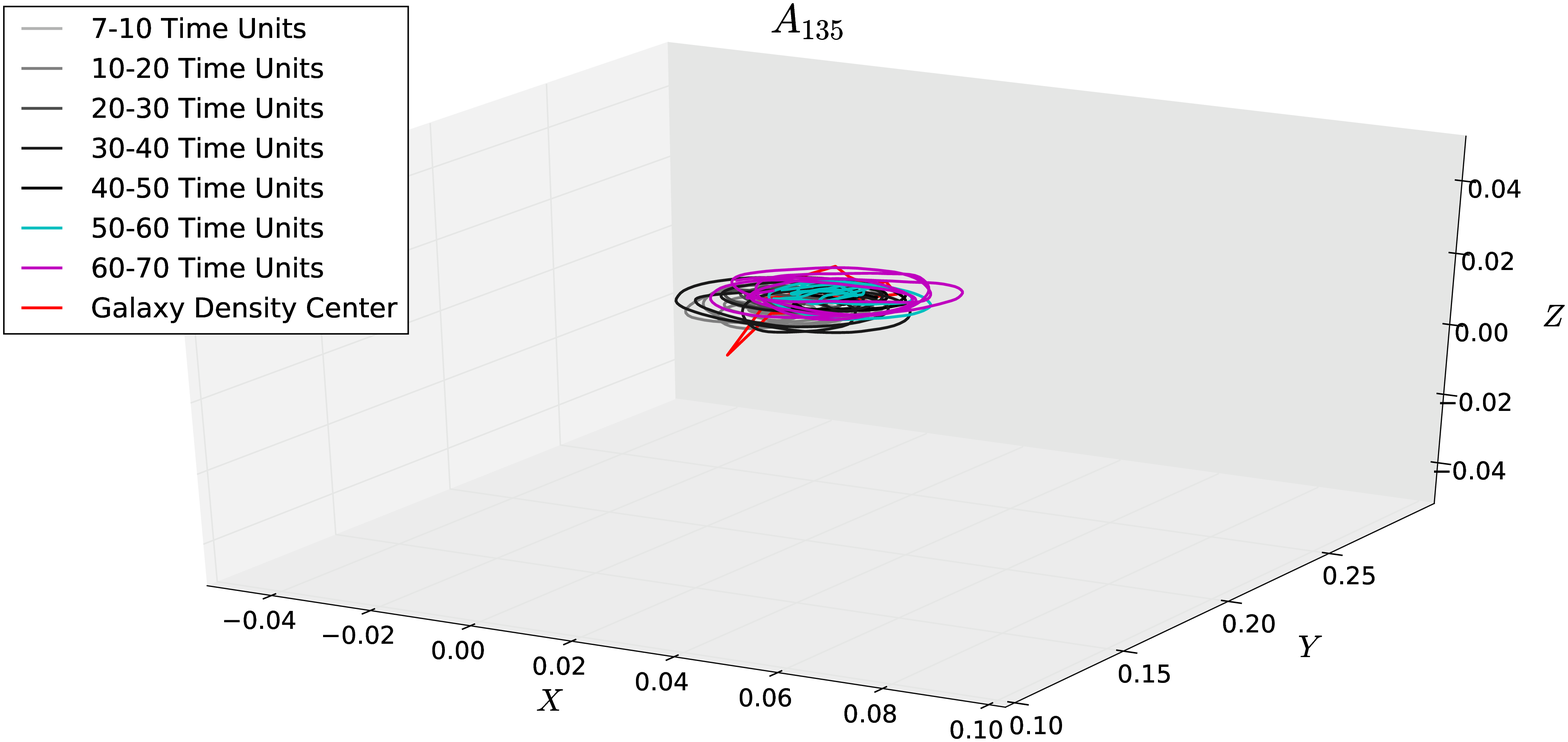}
		\includegraphics[width=1\linewidth, height = 1.7in]{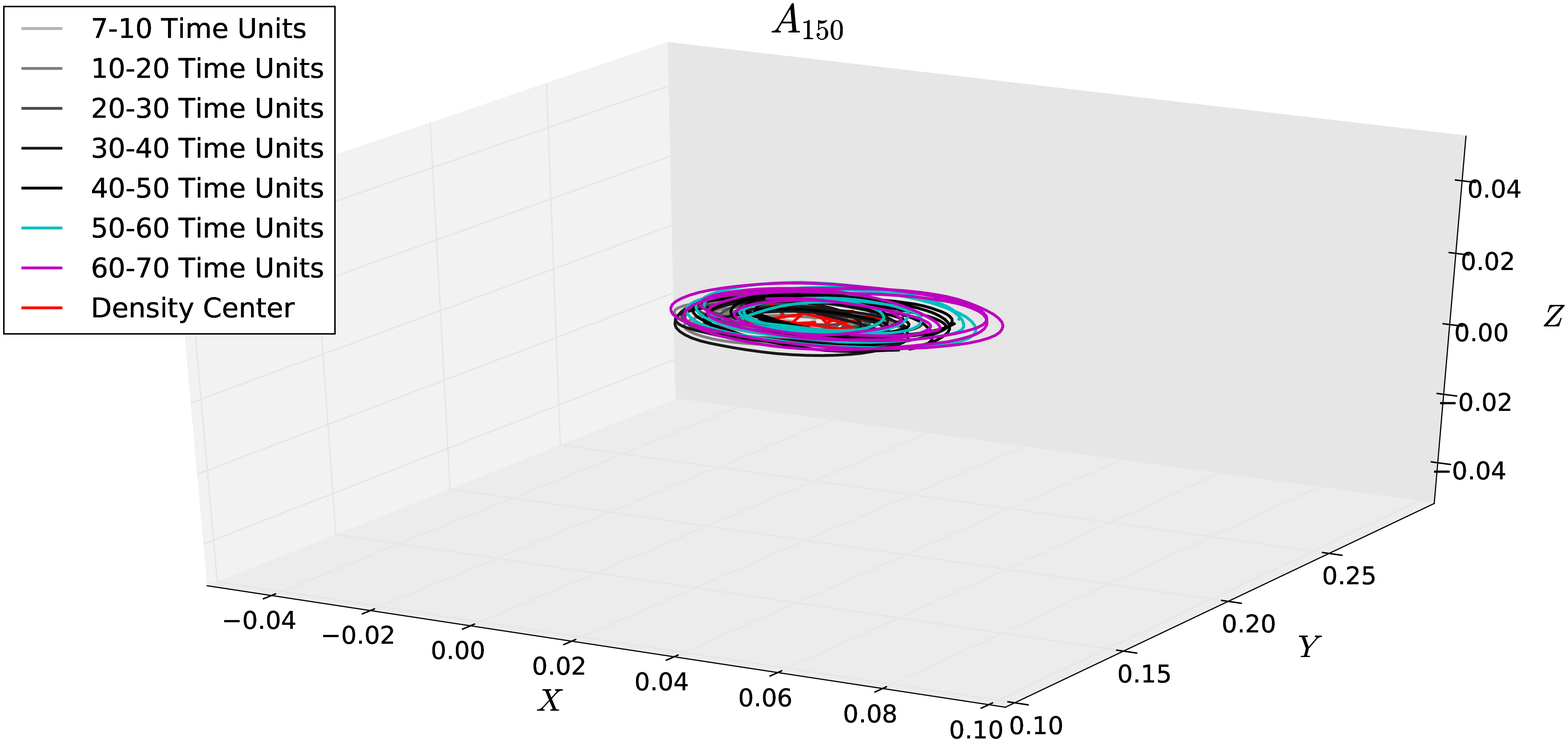}   	
		
		\caption{Center of mass motion of the SMBH binary within the galactic center for  A models. From top to bottom: $A_{45}$, $A_{90}$, $A_{120}$, $A_{135}$ and $A_{150}$. The legend lists the colors used to distinguish the trajectory at different time intervals.  Notice the wide extent of the orbit for  $A_{45}$,  $A_{90}$ and  $A_{120}$. }
		
		\label{fig:comA}
	\end{figure}
	
	\begin{figure}
		\centering
		\includegraphics[width=1\linewidth]{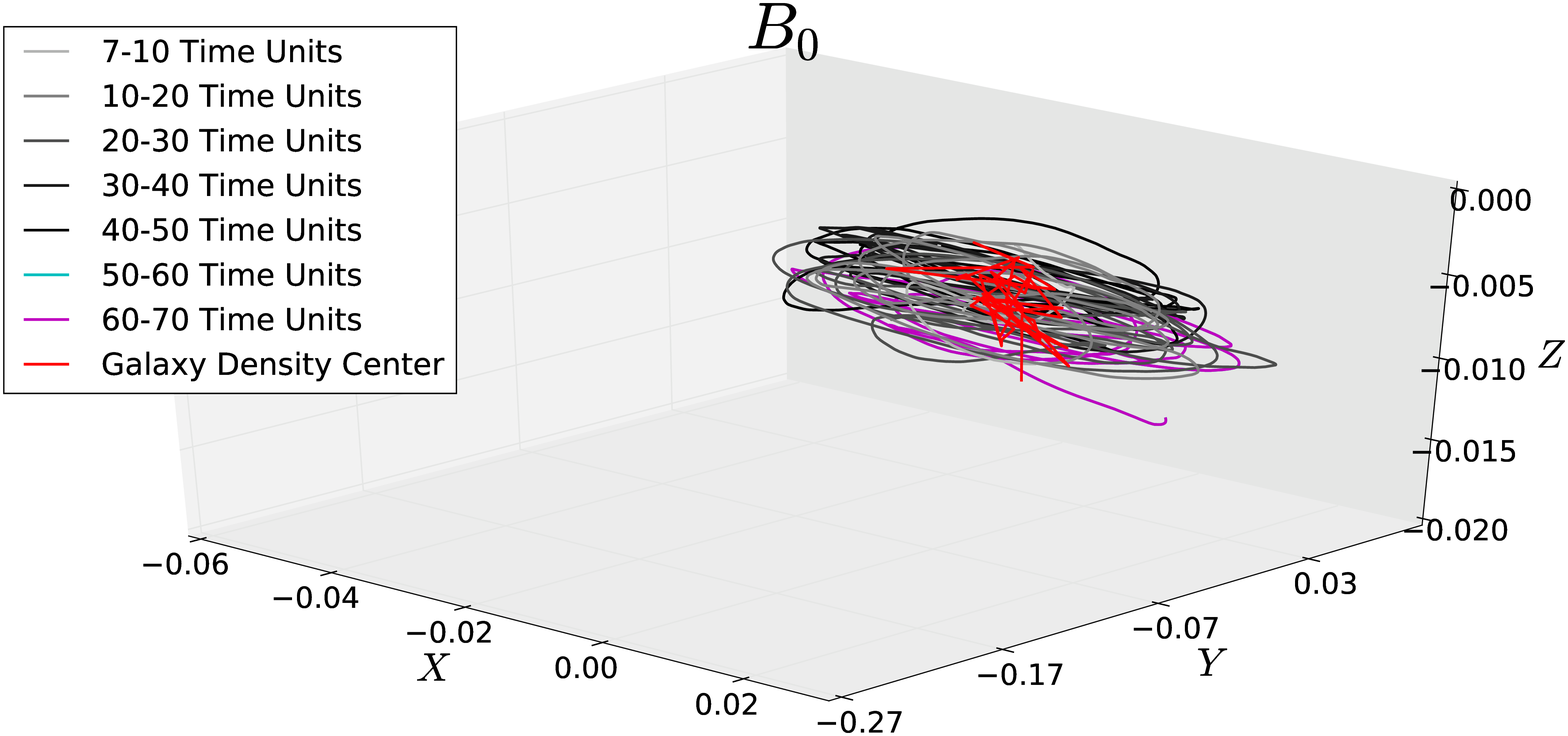}   	
		\includegraphics[width=1\linewidth]{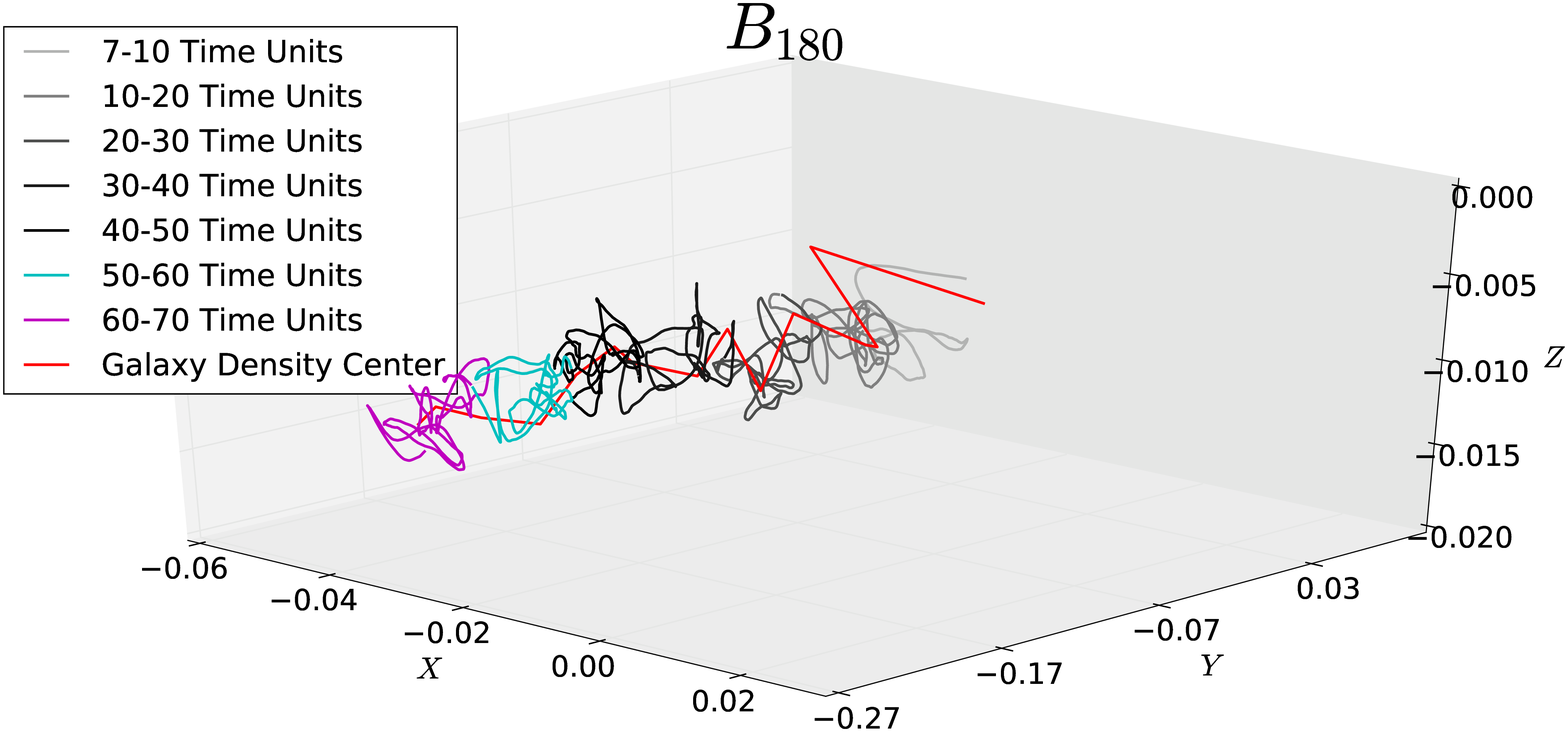}
		\caption{Center of mass motion of the SMBH binary in partially rotating B models. Upper panel: $B_0$, Lower panel: $B_{180}$. Here the center of mass is observed to neatly follow the trajectory of the density center; note that the axis limits here are zoomed in compared to figure 8. }
		\label{fig:comB}
	\end{figure}

	The behavior of the center of mass of the binary in our A models is shown in figure \ref{fig:comA} . We observe that for $A_{45}$, $A_{90}$ and $A_{120}$, the binary center of mass orbit spans a wide spatial extent; the orbit is in the range of 0.10 times the scale radius, which is of the order of the influence radius of the SMBH binary $($0.085 scale radius$)$. This is a behavior that has not been found in previous work on non-rotating \citep{Chatterjee+02} or counterrotating models (HB15). As the binary reaches a depth in the potential where the center of mass and galactic velocities are commensurate, dynamical friction becomes inefficient in dragging the binary to the galactic center. When not limited to the precise center of the potential,  these SMBH binaries have interaction with more stars. Bearing this out, the binaries in the $A_{135}$ and $A_{150}$ models traverse a much more limited spatial extent.

This may indicate that SMBH binaries can be found far from the galactic center, and therefore future SMBH binary surveys that concentrate only on the precise galactic center may miss an important population. Note, though, that the center of mass of the galaxy also shifts to follow the SMBH binary once the binary delves deep enough into the galaxy to comprise a significant fraction of the mass.

	Figure  \ref{fig:comB} reveals the trajectory of the center of mass of both the binary and galaxy density center  (the center of mass of the stellar particles as determined by a shrinking-sphere method.) for our B models. In the $B_0$ case, the center of mass of the binary orbits the density center, while in the $B_{180}$ scenario, both the binary center of mass and galaxy density center exhibits a random walk.
    
    \subsection{SMBH merger coalescence times}

Taking the prior results into account, Table \ref{tab:final} calculates the SMBH binary merger timescale in the A and B runs; this specifically includes the time to form the binary, pass through the 3-body scattering and gravitational radiation phases, and merge the black hole binary. We present coalescence times for two SMBH masses, namely, $10^6 M_{\sun}$ and $10^8 M_{\sun}$, in Table \ref{tab:final}.

For $10^6 M_{\sun}$ black hole scaling, our merger timescales range from 40 - 500 Myr, and for $10^8 M_{\sun}$ black hole scaling, the merger timescales range between 100 Myr to 1 Gyr. While these timescales are much less than a Hubble time, our longest coalescence would occur well after the galaxy merger -- a galaxy merger at redshift 6, for example, would be followed by a SMBH merger at redshift 3, and after such a delay, the galaxy merger signature may be undetectable. Our main point is that black hole mergers may follow galaxy mergers with a observationally significant delay, but we caution that the particular example at redshift 6 could be misleading simply because high redshift galaxies tend to be much more compact than the present-day galaxies. Black hole coalescence timescales in high redshift galaxies may in fact be very short \citep{Khan+16}; this emphasizes the need for more complete studies of the interplay between SMBH and host galaxy dynamics using more realistic observationally-based models.

\begin{table*}{}
	\centering
	\caption{Table of final parameters and coalescence times.\label{tab:final}}
		\begin{tabular}{ccccccc}
\hline
	Model & s & $e_{f}$ & $a_{f}$ & $T_{f}$ & $T_{coal,M_{1}}$  & $T_{coal,M_{2}}$\\
\hline
	$A_0$ & 27.7 & 0.17 & $3.29\times10^{-4}$ & 100 & (1901, 525 Myr) & (687, 920 Myr)\\ 
	$A_{45}$ & 33.9 & 0.04 & $3.87\times10^{-4}$ & 70 & (1650, 455 Myr) & (598, 801 Myr)\\ 
	$A_{90}$ & 32.2 & 0.05 & $3.56\times10^{-4}$ & 80 & (1718, 474 Myr) & (622, 833 Myr) \\ 
	$A_{120}$ & 33.5 & 0.59 & $3.72\times10^{-4}$ & 70 & (1250, 345 Myr) & (450, 603 Myr) \\ 
	$A_{135}$ & 26.15 & 0.72 & $4.06\times10^{-4}$ & 70 & (1239, 342 Myr) & (436, 584 Myr) \\ 
	$A_{150}$ & 29.31 & 0.93 & $4.09\times10^{-4}$ & 70 & (497, 137 Myr) & (171, 229 Myr) \\
	$A_{180}$ & 23.6 & 0.99 & $4.24\times10^{-4}$ & 73 & (140, 38 Myr) & (78, 104 Myr)\\ 
	$B_0$ & 30.5 & 0.14 & $4.19\times10^{-4}$ & 70 & (1773, 489 Myr) & (641, 859 Myr)\\
	$B_{180}$ & 21.65 & 0.92 & $5.03\times10^{-4}$ & 70 & (685, 189 Myr) & (235, 315 Myr)\\
	\end{tabular}
	
Column 1: Galaxy model.  Column 2: Hardening rate from 40 to 60 time units except A$_0$ and A$_{180}$ for which it is from 50 to 70 time units. Column 3: Eccentricity at $T_{f}$, the time when simulation was stopped. Column 4: Semi-major axis at $T_{f}$. Column 5: Time when simulation was stopped. Column 6: Coalescence time for $M_1 = 10^6 M_{\sun}$. Column 7: Coalescence time for $M_2 = 10^8 M_{\sun}$. In column 6 and 7, the first number is in model units while second number is in Myr.
\end{table*}

\section{Summary and Discussion}
	
	We examined how SMBH binaries are influenced by the surrounding rotation of galaxy, concentrating on three themes: 1) varying the angle of the SMBH binary orbital plane; 2) the effect of core rotation; 3) varying the SMBH initial orbital eccentricities. In counterrotating models, we found that eccentricity of the SMBH binary is highly dependent on the projection of the binary angular momentum vector onto that of the galaxy. Moreover, the C models showed that eccentricity increases for counterrotating models, with the largest eccentricity changes occurring for orbits with moderately high eccentricity in the first place.Torque is strong on models with an inclination that subjects the binary to counterrotating stars such that the binary plane flips to a more corotating orientation. From this, we may expect SMBH binaries in real galaxies to be predominately co-rotating. 
	
	We find that the central galaxy kinematics are primarily responsible for the characteristic behavior of the binary, and anticipate that a kinematic core a few times the binary mass is needed to drive the SMBH coalescence.  However, we find that the center of mass of a SMBH binary can fall into an orbit well outside the density center of a galaxy, which has observational consequences that may complicate dual-AGN surveys.

 	For hosts of LISA-like SMBHs, our merger timescales range between 40 Myr to 500 Myr; while these timescales are much less than a Hubble time, our longest coalescence timescale would occur long after the galaxy merger. This could mean that tying SMBH merger rates to observationally-derived galaxy merger rates will require a non-trivial statistical correction for SMBH coalescence timescales, and could complicate campaigns to determine AGN duty cycles. Theoretical understanding of SMBH demographics from hydrodynamic simulations and semi-analytic models will need to take these relatively long coalescence timescales into account when activating AGN feedback and SMBH growth. One bright consequence, however, is that this delay in merger timescales may be a boon to gravitational wave observatories such as LISA in that the eventual SMBH mergers will be louder sources. 
    
    It is increasingly clear that the galaxy kinematics, shape, structure, and gas content all matter in determining the evolution of a SMBH binary. From this work, it is evident that the details of the secondary SMBH orbit matter too. Differences in the initial orbit inclination and eccentricity alone can introduce a variation in SMBH evolution merger times for a given (relatively simple) galaxy host that spans an order of magnitude. Eccentricities in the gravitational wave regime are highly dependent on the initial conditions as well, though most generic configurations seem to result in nearly radial SMBH binaries. To prepare to estimate SMBH binary parameters from LISA observations, it may be beneficial to construct a large suite of waveforms with significant eccentricity, though note that the burst of gravitational radiation near pericenter may act to circularize the orbit again.
    
     In addition to the prior work in this area discussed in the Introduction and Results section, we note that at the time of writing this manuscript we came across an unpublished study by \cite{Rasskazov+16} which addresses overlapping concerns using a Fokker-Planck technique with numerically-informed drift and diffusion coefficients that solve for the orientation of the binary orbital plane within a rotating and non-evolving spherical nucleus. Despite the very different methodology, their broad findings are consistent with our own and those of prior work, in that the angular momenta of the nuclear and the binary SMBH tend to seek alignment, that the eccentricity increases for counterrotating systems, and that the eccentricity and inclination evolve in tandem. 

\section*{Acknowledgments}

	We acknowledge the support by Vanderbilt university for providing the access to its Advanced Computing Center for Research and Education (ACCRE). 
    
    FK acknowledges the support of Higher Education Commission of Pakistan through National Research Program for Universities(NRPU) project 4159.
           
    PB acknowledge the support by Chinese Academy of Sciences through the
Silk Road Project at NAOC, through the `Qianren' special foreign experts
program and the President's International Fellowship for Visiting
Scientists program of CAS and also the Strategic Priority Research Program
(Pilot B) `Multi-wavelength gravitational wave universe' of the Chinese Academy
of Sciences (No. XDB23040100).

PB acknowledge the support of the Volkswagen Foundation under the Trilateral
Partnerships grant No. 90411 and the special support by the NASU under
the Main Astronomical Observatory GRID/GPU computing cluster project.

Farrukh Chishtie would like to thank the hospitality of the CERN Theory group where part of this work was conducted.



\bibliographystyle{mnras}
\bibliography{ms}

\bsp	
\label{lastpage}
\end{document}